\documentclass[article]{aa}
\usepackage{epsf}
\usepackage{times}

\newcommand{\msun}{$\rm M_{\sun}$}
\newcommand{\lsun}{$\rm L_{\sun}$}
\newcommand{\teff}{$ T_{\rm eff}$}

\begin{document}
\topmargin0.0cm

   \thesaurus{06         % A&A Section 6: Form. struct. and evolut. of stars
              (08.05.3;
               08.06.3;
               08.09.3;
               08.23.1)
             }
  \title{Mass - radius relations for white dwarf stars of
different internal compositions}

\author{J. A. Panei\thanks{Fellow of the Universidad Nacional de
La Plata, Argentina.}, L. G. Althaus\thanks{Fellow of the
Consejo Nacional de Investigaciones Cient\'{\i}ficas y T\'ecnicas
(CONICET),  Argentina.},  O.  G.  Benvenuto\thanks{Member of the
Carrera   del   Investigador   Cient\'{\i}fico,   Comisi\'on   de
Investigaciones Cient\'{\i}ficas de la Provincia de Buenos Aires,
Argentina.}
}
\offprints{J. A. Panei}

\institute{
Facultad  de  Ciencias
Astron\'omicas  y  Geof\'{\i}sicas,
Paseo del Bosque S/N, (1900) La Plata, Argentina \\
panei,althaus,obenvenuto@fcaglp.fcaglp.unlp.edu.ar
}

   \date{Received; accepted }

\maketitle

\markboth{J.A.Panei et al. Mass-radius relations for white
dwarfs}{}

\begin{abstract} The purpose of this work is to present  accurate
and detailed mass - radius relations for white dwarf (WD)  models
with helium, carbon, oxygen, silicon and iron cores, and with and
without a  hydrogen envelope,  by using  a fully  updated stellar
evolutionary code. We considered  masses from 0.15 \msun\  to 0.5
\msun\  for  the  case  of  helium  core, from 0.45 \msun\ to 1.2
\msun\ for carbon, oxygen and silicon cores, and from 0.45 \msun\
to 1.0 \msun\ for the case of an iron core. In view of the recent
measurements made  by {\it  Hipparcos} that  strongly suggest the
existence of  WDs with  an iron  - dominated  core, we  focus our
attention  mainly  on  the  finite  -  temperature, mass - radius
relations  for  WD  models  with  iron interiors. In addition, we
explore  the  effects  of  gravitational,  chemical  and  thermal
diffusion on low mass helium white dwarf models with hydrogen and
helium envelopes.

      \keywords{stars: fundamental parameters -- stars: white
               dwarfs -- stars: evolution -- stars: interiors
               }
   \end{abstract}

\section{Introduction}

It is a well known fact  that about 90\% of stars will  end their
lives as  white dwarf  (WD) stars.  At present  we know different
routes that drive  stellar objects to  such a fate.  It is widely
accepted, for instance, that low mass WDs with stellar masses  $M
\la 0.45$ \msun\  are composed of  helium and that  they have had
time  enough  to  evolve  to  such  state  as a result of binary
evolution. For  intermediate mass  WDs, stellar  evolution theory
predicts an internal composition dominated by carbon and  oxygen.
Finally, for  the high  mass tail  of the  WD mass  distribution,
theory predicts interiors made up by neon and magnesium.

Over the  years, it  has been  customary to  employ mass - radius
relations  to  confront  theoretical  predictions on the internal
composition of WDs with  observational data. This is  so because,
as it is  well known since  Hamada \& Salpeter  (1961) (hereafter
HS,  see  also  Shapiro  \&  Teukolsky  1983), zero - temperature
configurations are sensitive to the internal composition. One  of
the  effects  that  allow  us  to  discriminate  the  WD internal
composition for a given stellar mass is related to the dependence
of the non - ideal  contributions to the equation of  state (EOS)
of degenerate matter (such  as Coulomb interactions and  Thomas -
Fermi   corrections)   on   the   chemical   composition.   These
contributions to the EOS are larger the higher the atomic  number
$Z$ of the chemical constituent. Another very important effect is
that, in  the case  of heavy  elements like  iron, nuclei  are no
longer  symmetric  ($Z=  26,  A=  56$  for iron), yielding a mean
molecular weight per electron  higher than 2. Accordingly,  for a
fixed mass value, the WD radius is a decreasing function of $Z$.

Recently, Provencal  et al.  (1998) (and  other references  cited
therein)  have  presented  the  {\it  Hipparcos} parallaxes for a
handful of  WDs. These  parallaxes have  enabled to significantly
improve  the  mass  and  radius  determination  of some WDs, thus
allowing for a  direct confrontation with  the predictions of  WD
theory. In particular, the suspicion that some WDs would fall  on
the zero -  temperature, mass -  radius relation consistent  with
iron cores (see Koester \&  Chanmugam 1990) has been placed  on a
firm observational ground by these satellite - based measurements
(see Provencal et  al. 1998) (however,  see below). Indeed,  some
WDs have much smaller radii than expected if their interior  were
made of carbon and oxygen, suggesting that, at least, two of  the
observed WDs have  iron - rich  cores. Specifically, the  present
determinations indicate that Procyon B  and EG 50 have radii  and
masses consistent with zero  - temperature, iron WDs.  Obviously,
such  results  are  in  strong  contradiction  with  the standard
predictions of stellar evolutionary calculations, which allow for
an iron - rich interior only in the case of presupernova objects.
Although  these  conclusions   are  based  on   the  HS  zero   -
temperature, mass  - radius  relations (note  that EG  50 has  an
effective temperature, \teff, of \teff $\approx 21000$ K), it  is
clear that unless observational determinations are incorrect, the
interior  of  the  above  -  mentioned  WDs  is  much denser than
expected before.

Before the above - mentioned determinations, an iron  composition
has  been  considered  as  quite  unexpected.  In  fact, the only
attempt of proposing a physical  process able to account for  the
formation of iron WDs is, to our knowledge, that of Isern et  al.
(1991).  In  their  calculations,  Isern  et  al.  find  that  an
explosive  ignition  of  electron  -  degenerate ONeMg cores may,
depending critically upon the  ignition density and the  velocity
of  the  burning  front,  give  rise  to the formation of neutron
stars, thermonuclear supernovae or iron WDs. It is therefore  not
surprising that,  apart from  the study  carried out  long ago by
Savedoff  et  al.  (1969),  who  did  not consider the effects of
electrostatic  corrections,  convection  and  crystallization  in
their calculations, very  little attention has  been paid to  the
study of the evolution of iron WDs.

We should warn the reader that the existence of WDs with an  iron
- rich  interior is  still under  debate. In  particular, despite
recent claims of an  iron - rich interior  for Procyon B, in  the
report  of  this  work  our  referee  has  told  us  about  a new
reanalisis  of  the  observational  data  which,  in a preliminar
stage, seems to indicate an interior composition for this  object
consistent with a carbon one.

Another interesting possibility is that these objects may content
some extremely compact  core, as proposed  by Glendenning et  al.
(1995a,  b).  They  suggested  the  existence  of stellar objects
composed by a strange quark matter (with a density of $\approx  5
\times 10^{14}$\ g\ cm$^{-3}$) surrounded by an extended,  normal
matter envelope. These configurations have been called  ``strange
dwarfs''. It is  presently known that  these objects have,  for a
given mass and chemical composition for the normal matter layers,
a much lower radius than a standard WD, and also that they evolve
in a  very similar  way compared  to standard  WDs (Benvenuto  \&
Althaus  1996a,  b).  However,  at  present,  it  is difficult to
account for the formation of a strange quark matter core inside a
WD star.

In view of the above  considerations, we present in this  paper a
detailed  set  of  mass  -  radius  relations  for WD models with
different assumed internal compositions, with the emphasis placed
on models  with iron  - rich  composition. Despite  the fact that
many researchers have addressed the problem of theoretical mass -
radius relations for WD of helium (Vennes et al. 1995;  Benvenuto
\& Althaus 1998; Hansen \& Phinney 1998 and Driebe et al.  1998),
carbon  and  oxygen  (Koester  \&  Sch\"onberner  1986; Wood 1995
amongst  others),  we  judge  it  to  be worthwhile to extend our
computations to the case of models with these compositions in the
interests of presenting an homogeneous sequence of mass -  radius
relations. In particular, we  shall consider the internal  layers
as  made  up  by  helium  ($^{4}$He),  carbon  ($^{12}$C), oxygen
($^{16}$O), silicon ($^{28}$Si) and iron ($^{56}$Fe),  surrounded
by a helium  layer with a  thickness of $10^{-2}\;  M_{*}$ (where
$M_{*}$  is  the  stellar  mass).  We  considered  models with an
outermost  hydrogen  layer   of  $10^{-5}\;  M_{*}$   ($3  \times
10^{-4}\; M_{*}$  in the  case of  helium core  models) and  also
models  without  any  hydrogen  envelope.  In  doing so, we shall
employ a full stellar evolution code, updated in order to compute
the properties of iron - rich, degenerate plasmas properly.

This paper is organized  as follows. In Sect.  \ref{sec:code}, we
present the general structure of  the computer code that we  have
employed and the main improvements we have incorporated in it. In
Sect. \ref{sec:results}, we describe the strategy employed in the
computations,  and  the  numerical  results.  Finally,  in  Sect.
\ref{sec:conclusion},  we  discuss  the  main  implications of
our results.

\section{The computer code} \label{sec:code}

The  WD  evolutionary  code  we  employed  in this study is fully
described in Althaus \& Benvenuto (1997, 1998), and we refer  the
reader to  those works  for a  general description.  Briefly, the
code is  based on  the technique  developed by  Kippenhahn et al.
(1967)  for  calculating  stellar  evolution,  and  it includes a
detailed  and  updated  constitutive  physics  appropriate  to WD
stars. In  particular, the  EOS for  the low  - density regime is
that of  Saumon et  al. (1995)  for hydrogen  and helium plasmas,
whilst the treatment for  the completely ionized, high  - density
regime  includes   ionic  contributions,   coulomb  interactions,
partially degenerate  electrons, electron  exchange and  Thomas -
Fermi contributions at  finite temperature. Radiative  opacitites
for  the  high  -  temperature  regime  ($T  \geq  6 000$ K) with
metallicity $Z=0$ are  those of OPAL  (Iglesias \& Rogers  1993),
whilst for lower  temperatures we use  the Alexander \&  Ferguson
(1994) molecular opacities.

High - density conductive opacities and the various mechanisms of
neutrinos emission for different chemical composition  ($^{4}$He,
$^{12}$C,  $^{16}$O,  $^{20}$Ne,  $^{24}$Mg, $^{28}$Si, $^{32}$S,
$^{40}$Ca and  $^{56}$Fe) are  taken from  the works  of Itoh and
collaborators (see  Althaus \&  Benvenuto 1997  for details).  In
addition   to   this,   we   include   conductive  opacities  and
Bremsstrahlung  neutrinos  for  the  crystalline  lattice   phase
following Itoh et  al. (1984a) and  Itoh et al.  (1984b; see also
erratum  Itoh  et  al.  1987),  respectively.  The latter becomes
relevant for WD models with iron core since these models begin to
develop a crystalline core at high luminosities (up to two orders
of  magnitude  higher  than  the  luminosity  at which a carbon -
oxygen WD of the same  mass begins to crystallize). With  respect
to  the  energy  transport   by  convection,  for  the   sake  of
simplicity,  we  adopt  the  mixing  length  prescription usually
employed in  most WD  studies. This  choice has  no effect on the
radius of the models. Finally, we consider the release of  latent
heat during crystallization  in the same  way as in  Benvenuto \&
Althaus (1997).

As in our previous works on WDs, we started the computations from
initial models at a far higher luminosity than that corresponding
to  the  most  luminous  models  considered as meaningful in this
paper. The procedure we follow to construct the initial models of
different  stellar  masses  and  internal chemical composition is
based on  an artificial  evolutionary procedure  described in our
previous papers cited above.  In particular, to produce  luminous
enough  initial  models,  we  considered  an  artificial   energy
release. After such ``heating'', models experience a  transitory
relaxation to  the desired  WD structure.  Obviously, the initial
evolution of our WD models is affected by this procedure but, for
the  range  of  luminosity  and  \teff\ values considered in this
paper  this  is  no  longer  relevant  (see below) and our mass -
radius relations are completely meaningful.

\section{Numerical results} \label{sec:results}

In order to compute accurate mass - radius relations, we  evolved
WD models with masses ranging  from 0.15 \msun\ to 0.5  \msun\ at
intervals of  5\% for  helium core  WDs; from  0.45 \msun\ to 1.2
\msun\ at intervals of 0.01 \msun\ for carbon, oxygen and silicon
cores; and finally from 0.45 \msun\ to 1.0 \msun\ for the case of
an  iron  core  at  intervals  of  0.01  \msun.  The evolutionary
sequences were computed down to $\log L$/\lsun= -5. Mass - radius
relations  for  interior  compositions  of  $^{12}$C,   $^{16}$O,
$^{28}$Si and $^{56}$Fe are  presented for \teff\ values  ranging
from \teff= 5000K to 55000 K with steps of 10000 K and from 70000
to 145000  K with  steps of  15000 K.  For the  case of helium WD
models we considered \teff\ values from \teff= 4000 K to 20000  K
with steps of 4000 K.  To explore the sensitivity of  our results
to a hydrogen envelope, we considered two values: $M_{\rm  H}/M_*
= 10^{-5}$  ($3 \times  10^{-4}\; M_{*}$  in the  case of He core
models) and $M_{\rm H}/M_* = 0$. For the sake of comparison,  for
each of the  considered core compositions  we have also  computed
the zero - temperature HS models.

It is worth  noting that all  the models included  in the present
work have densities below  the neutronization threshold for  each
chemical  composition  ($1.37  \times  10^{11}$  g\ cm$^{-3}$ for
helium,  $3.90  \times  10^{10}$  g\  cm$^{-3}$ for carbon, $1.90
\times 10^{10}$ g\ cm$^{-3}$ for oxygen, $1.97 \times 10^{9}$  g\
cm$^{-3}$ for silicon, and $1.14 \times 10^{9}$ g\ cm$^{-3}$  for
iron).  Such  densities  represent  the  end  of the WD sequences
because  electron  capture  softens  the  EOS  and  the   stellar
structure becomes  unstable against  gravitational collapse  (see
Shapiro \& Teukolsky 1983 for further details).

In recent  years, both  observational (Marsh  1995; Moran  et al.
1997; Landsman et al. 1997;  Edmonds et al. 1999 amongst  others)
and theoretical (Althaus \& Benvenuto 1997, Benvenuto \&  Althaus
1998, Hansen \&  Phinney 1998, Driebe  et al. 1998)  efforts have
been devoted to the study of helium WDs. It is now accepted  that
these objects  would be  the result  of the  evolution of certain
binary systems, in which mass transfer episodes would lead to the
formation of helium degenerates within a Hubble time (see,  e.g.,
Iben \& Tutukov 1986; Alberts  et al 1996; Ergma \&  Sarna 1996).
Connected with the age determination for millisecond pulsars from
WD cooling is the existence or not of hydrogen flashes in  helium
WDs. In particular,  detailed calculations predict  that hydrogen
flashes do  not occur  on WDs  of mass  less than 0.2 $M_{\odot}$
(see also Driebe et al. 1998), but instead such low - mass helium
degenerates experience long - lasting phases of hydrogen  burning
(but see Sarna et al. 1998).

With regard to the  main topic of the  present work, it is  worth
mentioning that Vennes, Fontaine  \& Brassard (1995) presented  a
set of static  mass - radius  relations for hot  WDs However, the
authors  considered  a  linear  relation  between  the   internal
luminosity  and  the  mass,  thus  avoiding  the  computation  of
evolutionary  sequences.  This  approximation  is  equivalent  to
neglecting neutrino emission, which is not a good assumption  for
their hottest models.

In  a  recent  paper,  Driebe  et  al.  (1998)  have computed the
evolution of  low mass  stars from  the main  sequence up  to the
stage of helium  WD. In that  work the binary  evolution has been
mimicked by applying, at  appropriate positions, large mass  loss
rates  from  a  single  star.  More  importantly,  diffusion  was
neglected throughout  the entire  evolution. In  this connection,
gravitationally  induced  diffusion   is  expected  to   lead  to
noticeable  changes  in  the  surface  gravity of their helium WD
models, the envelope of which at the end of mass loss phase is  a
mixture of helium and  hydrogen. Indeed, during their  evolution,
WDs should  modify the  outer layers  chemical composition making
essentially the bulk of the hydrogen float to the surface and the
helium  sink  out  of  surface  layers.  In this way, this effect
causes  the  outer  layers   composition  to  approach  to   pure
composition layers,  the case  we assumed  in the  present paper.
Preliminary results to be presented below indicate this to be the
actual case,  as we  suggested previously  (Benvenuto \&  Althaus
1999).
\begin{figure}[h]
\epsfxsize=8.8cm
\epsfbox{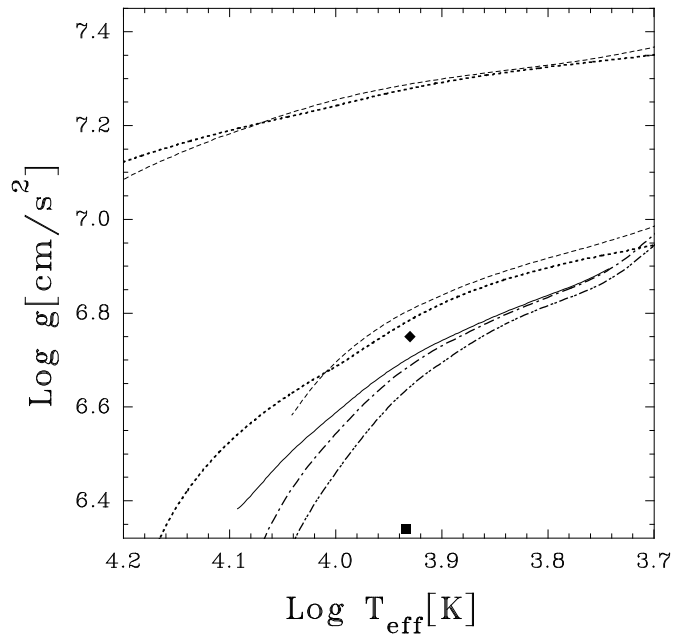}
\caption{
Surface gravity versus \teff for
0.3 (upper curves) and 0.195 (lower curves) $M_{\odot}$ helium WD
models. Dotted lines correspond to models calculated by Driebe et
al. (1998). Dashed  lines represent our  models with an  envelope
mass of 1.2 $\times  10^{-3} M_{\odot}$ (with a  hydrogen content
by mass of $X_{\rm H}$=  0.538) and 6 $\times 10^{-4}  M_{\odot}$
(with $X_{\rm H}$= 0.7) for the 0.195 and 0.3 $M_{\odot}$ models,
respectively. Solid line corresponds  to the case when  diffusion
is included in  our 0.195 $M_{\odot}$  model with an  envelope of
1.2  $\times  10^{-3}  M_{\odot}$  and  an  initial $X_{\rm H}$=
0.538, while dot - dashed lines and dot - dot - dashed lines  are
for our  0.195 $M_{\odot}$  models with  pure hydrogen  envelopes
with  mass  of  $6  \times  10^{-4}$  and  $1.2  \times  10^{-3}$
$M_{\odot}$, respectively.  The location  of the  WD companion to
the millisecond pulsar PSR J1012 + 5307 according to van Kerkwijk
et al. (1996)  and Callanan et  al. (1998) determinations  (upper
and lower square, respectively) are also indicated.}
\end{figure}

To  address  the  problem  of  diffusion  in  helium WDs, we have
developed   a   code   which   solves  the  equations  describing
gravitational settling and chemical and thermal diffusion.  Here,
we  present  some  details  of  our  code,  deferring  a thorough
description to a further  publication. In broad outline,  we have
solved the diffusion and heat flow equations presented by Burgers
(1969) for the  case of a  multicomponent medium appropriate  for
the case  we are  studying here  (see also  Muchmore 1984  for an
application of  the set  of Burgers's  equations to  the study of
diffusion in WDs). The resistence coefficients are from  Paquette
et  al.  (1986).  To  solve  the  continuity  equation  we   have
generalized  the  semi   -  implicit  finite   difference  method
presented by Iben  \& McDonald (1985)  to include the  effects of
thermal diffusion. We have followed the evolution of the isotopes
$^{1}$H, $^{3}$He, $^{4}$He, $^{12}$C and $^{16}$O. The diffusion
code has  been coupled  to our  evolutionary code  to follow  the
chemical evolution of our models self - consistently.
\begin{figure}[h]
\epsfxsize=8.8cm
\epsfbox{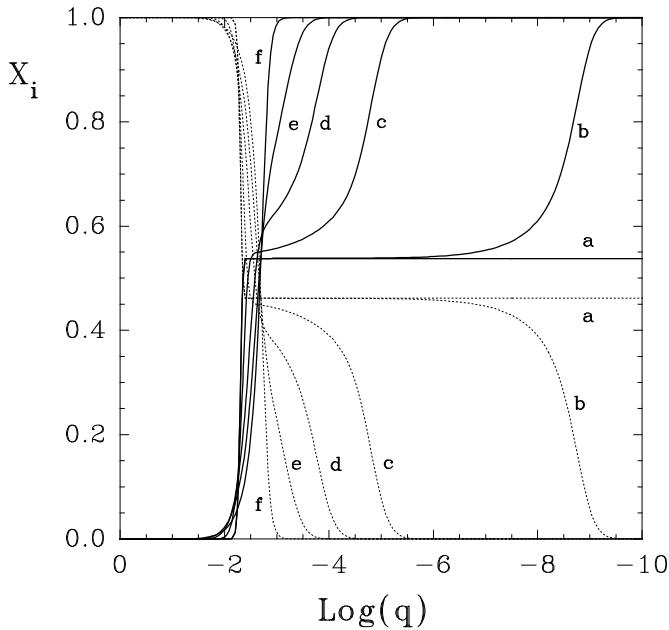}
\caption{
Evolution of the hydrogen (solid
lines) and helium  (dotted lines) profiles  as a function  of the
outer  mass  fraction  $q$  for  the  0.195 $M_{\odot}$ helium WD
model. Starting  out from  a model  with an  initially homogenous
envelope with $X_{\rm H}$= 0.538 (curves a), following models (b,
c, d, e and f) correspond to evolutionary stages characterized by
log \teff= 4.096, 4.063, 4.0, 3.932 and 3.762. }
\end{figure}

Let  us  first  compare  our  models  with those of Driebe et al.
(1998) in the case when diffusion is neglected. In Fig. 1 we show
the  surface  gravity  in  terms  of  \teff  for  0.195  and  0.3
$M_{\odot}$  helium  WD  models.  In  order  to  make  a   direct
comparison with Driebe et al.'s predictions, we have adopted  for
these  models  the  same  envelope  mass  and  hydrogen   surface
abundance as  quoted by  these authors.  The initial  models were
generated in  the same  fashion as  described previously. Despite
the assertions  by Driebe  et al.,  note that  our gravity values
after the  relaxation phase  of our  models are  very similar  to
those predicted  by these  authors. We  should remark  that their
``contracting models'' are very different from our initial  ones.
In fact,  they start  with a  homogeneous main  sequence model in
which nuclear energy release has been suppressed. Then, it is not
surprising  that  they  get  contracting  models  with  gravities
comparable to those obtained  with evolutionary models only  when
they  are  very  cool  (at  \teff  $\approx  $  3000K  for  a 0.2
$M_{\odot}$ model).  On the  contrary, in  our previous  works on
helium  WDs  with  hydrogen  envelopes,  we generated our initial
models from a  cool helium WD  model, adding to  it an artificial
energy  release  up  to  the  moment  in  which the model is very
luminous. Then, we switch it  off smoothly, getting a model  very
close to the cooling branch. Thus, notwithstanding Driebe et  al.
comments, our artificial  procedure gives rise  to mass -  radius
relations  in  good  agreement  with  those  found  with  a fully
evolutionary computation of the stages previous to the WD  phase.
A further comparison performed with  low - mass helium WD  models
calculated  by  Hansen  \&  Phinney  (1998)  with  thick hydrogen
envelopes  reinforces  our  assertion.  However, for more massive
models some divergences appear  between our results and  those of
Hansen \& Phinney.  Such differences are  the result of  the fact
that Hansen \& Phinney massive  models do not converge to  the HS
predictions for zero temperature configurations, a limit to which
our models tend.

Now, let us consider  what happens when diffusion  is considered.
To  this  end,  we  have  computed  the  evolution  of  a   0.195
$M_{\odot}$ helium WD model  with an envelope characterized  by a
mass  fraction  $M_{\rm  env}/M_{*}=  6  \times  10^{-3}$  and an
abundance by mass of hydrogen  ($X_{\rm H}$) of 0.538, as  quoted
in Driebe et al. (1998). We begin by examining Fig. 2 in which we
show  the  evolution  of  the  hydrogen  and helium profiles as a
function  of  the  internal  mass  fraction  for various selected
values  of  \teff.  Even  in  this  case  of low surface gravity,
diffusion proceeds in a very short timescale, giving rise to pure
hydrogen outermost layers.  If we start  out the computations  at
\teff  $\approx$   4.1,  we   find  that   at  the   \teff  value
corresponding to the WD companion to PSR J1012 + 5307, our  model
is  characterized  by  a  pure  hydrogen  envelope  of  $  M_{\rm
env}/M_{*} \approx 4 \times 10^{-4}$ (curve e). Needless to  say,
this will affect  the surface gravity  as compared with  the case
when  diffusion  is  neglected.  Thus,  in  order  to  accurately
estimate  the  mass  of  that  WD  we  do need to account for the
diffusion process.  This expectation  is borne  out by  Fig.1, in
which we have included the results corresponding to the situation
when diffusion is  included (solid line)  and to the  case of the
model with assumed pure hydrogen outer layers (dot dashed  lines)
throughout its  entire evolution  (i.e. the  conditions we  would
have  it  were  diffusion  instantaneous).  In both situations we
assumed that the  total initial amount  of hydrogen is  the same.
The differences in the value of the surface gravity compared with
the case of no diffusion  are noticeable. Finally, note that  the
track  asymptotically  merges   the  corresponding  to   complete
separation of hydrogen and  helium, the structures we  assumed in
our  previous   works.  These   results  clearly   justifies  the
assumptions we made in our  referred papers. We should also  note
that  in  the  case  of  a  somewhat  thicker  hydrogen envelope,
hydrogen  burning  increases   significantly,  thus  making   the
evolution considerably slower. Thus, in the plane surface gravity
versus \teff,  the asymptotic  conditions of  total separation of
hydrogen  and  helium  would  be  reached  far  earlier  in   the
evolution. Thus, diffusion is a fundamental ingredient if we want
a solid surface gravity versus \teff relation.
\begin{figure}[h]
\epsfxsize=8.8cm
\epsfbox{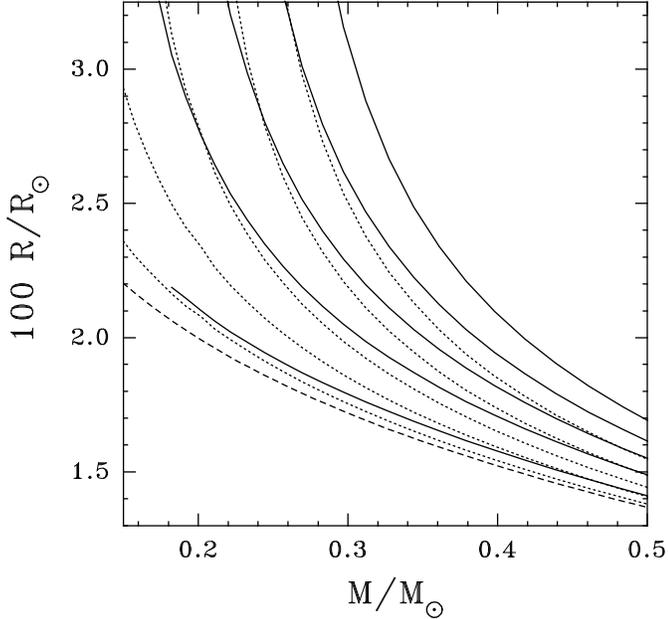}
\caption{
The mass - radius  relation for
WD  stars  with  a  helium  core.  Solid  and  short dashed lines
correspond  to  the  cases  of  objects  with  an  outermost pure
hydrogen  layer  of  $3  \times  10^{-4}\;  M_{*}$  and to models
without  hydrogen   layer,  respectively.   Medium  dashed   line
represent the mass  - radius relation  for HS homogeneous  helium
models. Finite temperature models are ordered from bottom to  top
with increasing \teff\ correspnding to  (in $10^{3}$ K) of 4,  8,
12, 16, and 20. }
\end{figure}

In Fig. 3,  we show the  mass - radius  relation for helium  core
models. We considered models with masses up to 0.5 \msun\ because
higher  mass  objects  should  be  able  to  ignite helium during
previous evolutionary stages  and should not  end their lives  as
helium WDs. As it is well known, models have a larger radius  the
higher the \teff. It is also clearly noticeable the effect on the
stellar  radius  induced  by  the  presence  of an outer hydrogen
envelope. These effects are  particularly important for low  mass
models. Let us consider the case of a 0.3 \msun\ helium WD model.
In  the  case  of  no  hydrogen  envelope,  at the highest \teff\
considered here,  the object  has a  radius $\approx$50\%  larger
than  that  corresponding  to  the  HS  model.  If we include the
hydrogen envelope, the radius is $\approx$80\% larger than the HS
one (let us remind the reader that in this case we have  included
a hydrogen layer 30  times more massive than  in the case of  the
other core compositions). Notice that for \teff $\rightarrow$  0,
the radius of the objects tends to the HS values, as expected.
\begin{figure}[h]
\epsfxsize=8.8cm
\epsfbox{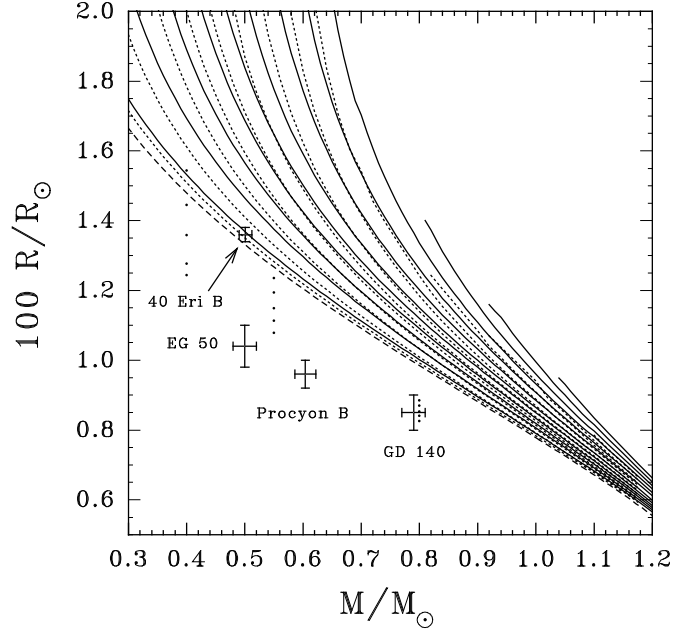}
\caption{
The mass - radius  relation for
WD stars with a carbon core  surrounded by a helium layer with  a
thickness of $10^{-2}\; M_{*}$. Solid and short dashed lines have
the  same  meaning  as  in  Fig.  3  but for case of the hydrogen
envelope we assumed  a mass of  $10^{-5}\; M_{*}$. Medium  dashed
line  corresponds   to  the   mass  -   radius  relationship  for
homogeneous HS carbon models. We have included in this figure the
values corresponding to \teff\ (in $10^{3}$ K) of 5, 15, 25,  35,
45, 55, 70, 85, 100, 115, 130 and 145. We have also included  the
data for strange dwarf models with \teff\ (in $10^{3}$ K) of  10,
20, 30, 40 and  50. Notice that they  are much more compact  than
standard WD  models with  the same  composition and  mass. In the
interests of completeness, we have extended the computations to a
mass value of 0.3 \msun\ . }
\end{figure}
\begin{figure}[h]
\epsfxsize=8.8cm
\epsfbox{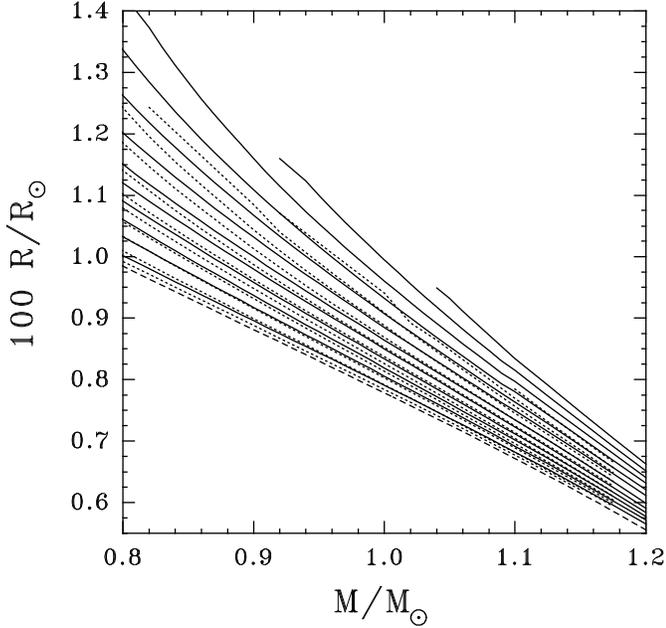}
\caption{
Same as Fig. 4, but for  the massive carbon core models.}
\end{figure}
In  Figs.  4  to  7  we  show  the results for carbon, oxygen and
silicon  interiors  respectively.  Although  the  effects  due to
finite temperature and the presence of an outer hydrogen envelope
are also noticeable, these are not so large as in the case of the
low  mass  helium  WD  models  shown  in the previous figure. For
example, for 1.2 \msun\ models, both effects are able to  inflate
the star only up to $\la 19 \%$. This is expected because as mass
increases,  internal  density  (and  electron  chemical potential
$\mu_{e}$) also increases. Thus, as thermal effects enter the EOS
of the degenerate gas as a correction $\propto  (T/\mu_{e})^{2}$,
EOS gets closer to the zero temperature behaviour, i.e. to the HS
structure. As  the thickness  of the  hydrogen layer  is $\propto
g^{-1}$ ($g$ is  the surface gravity)  it also tends  to zero for
very massive models. Note that carbon, oxygen and silicon have  a
mean  molecular  weight  per  electron  very  near  2 ($\mu_{e}=$
2.001299, 2.000000,  1.999364, and  1.998352 for  helium, carbon,
oxygen and silicon respectively), thus, the differences in  radii
for a given  stellar mass are  almost entirely due  to the non  -
ideal,  corrective  terms  of  the  EOS.  Because  of  this,  the
differences in  radii are  small, of  the order  of few percents.
Also, in each figure  we included the corresponding  HS sequence.
Notice that, for a given mass value, HS models have smaller radii
and that there exist some minute differences even for the  lowest
\teff\ models.  This is  due simply  to the  presence of a helium
(and hydrogen)  layer (if  present), the  effect of  which on the
model radius  was not  considered in  our computation  of the  HS
sequences.

\begin{figure}[h]
\epsfxsize=8.8cm
\epsfbox{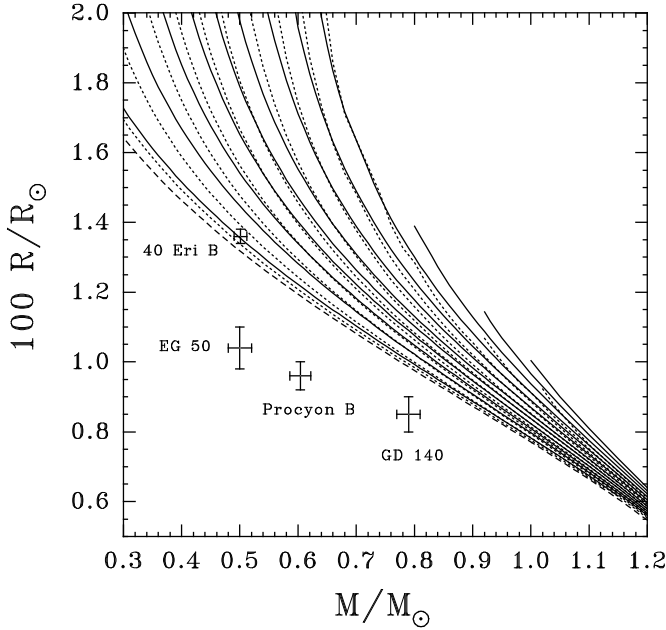}
\caption{
Same as Fig. 4, but for an oxygen core.}
\end{figure}
\begin{figure}[h]
\epsfxsize=8.8cm
\epsfbox{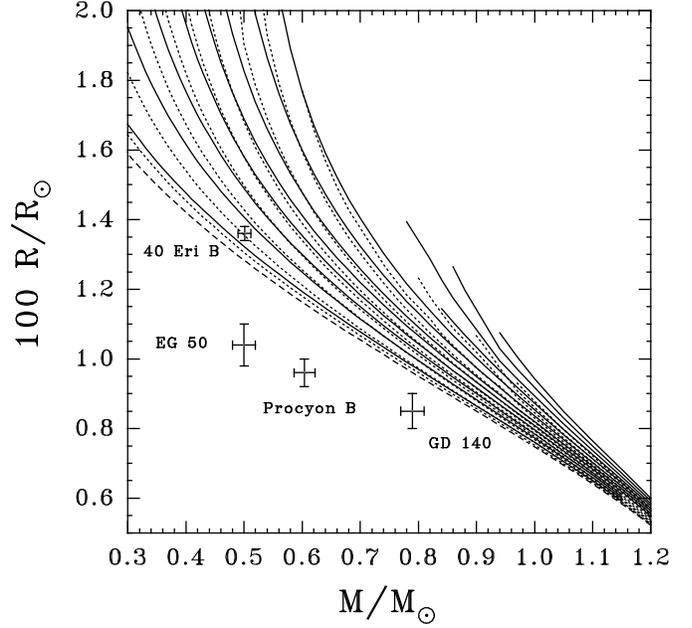}
\caption{
Same as Fig. 4, but for a silicon core.}
\end{figure}

In Fig. 4  we also included  the radii corresponding  to strange
dwarfs of 0.4, 0.55 and 0.8 \msun\ models for \teff\ from  \teff=
10000  K  to  50000  with  steps  of  10000 K. These objects were
computed assuming a  carbon - oxygen  composition for the  normal
matter envelope, but despite the precise chemical profile, it  is
clear that they have  much  smaller radii than WD models of  the
same mass.

In Fig. 8, we show  the mass - radius sequences  corresponding to
iron. These are noticeably  different from the previously  shown,
due  mainly  to  the  higher  mean  molecular weight per electron
($\mu_{e}= 2.151344$) and also  to the much higher  atomic number
($Z= 26$) that indicates a much strongly interacting,  degenerate
gas compared to the case of a standard composition.
\begin{figure}[h]
\epsfxsize=8.8cm
\epsfbox{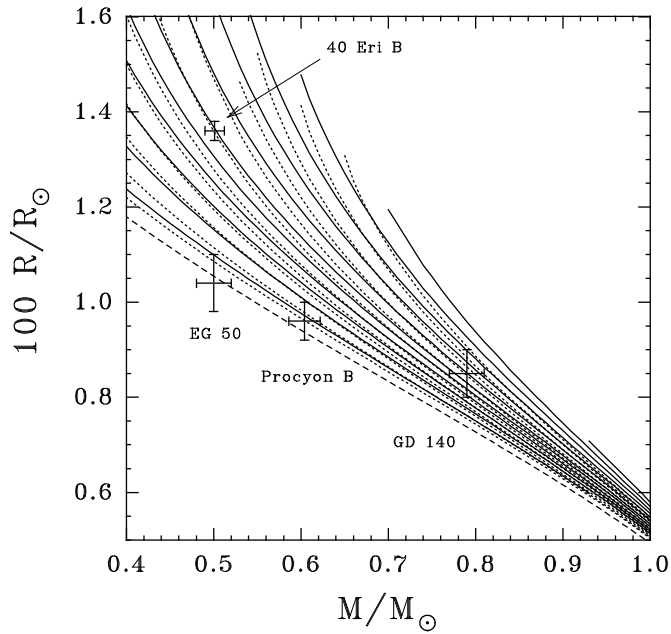}
\caption{
Same as  Fig. 4, but  for an
iron core. We have also included the data corresponding to 40 Eri
B, EG  50, Procyon  B and  GD 140  taken from  Provencal, et  al.
(1998). Note  the change  in the  vertical scale  compared to the
previous figures. For further discussion, see text.}
\end{figure}

In the case  of an iron  core, for a  fixed mass value,  the mean
density is almost  twice the corresponding  to carbon and  oxygen
cores. Thus, it is not surprising that, for the range of  \teff's
considered here, thermal effects  are less important than  in the
standard case.  For example,  for the  0.45 \msun\  iron model at
\teff $\approx$ 25000K, thermal  effects inflate its radius  only
by  $\approx  17\%$.  For  the  iron  core  case,  we  have  only
considered models up  to a mass  value of 1.0  \msun. Higher mass
objects are very near the  mass limit for such composition  (i.e.
the  central  density   becomes  very  near   the  neutronization
threshold, see  also Koester  \& Chanmugam  1990) and  would have
internal densities so high that our description of the EOS  would
not  be  accurate  enough  for  our  purposes.  The  evolutionary
sequences corresponding to an iron core composition are the  most
detailed and accurate computed to date, and a thorough discussion
of them will be deferred to a separate publication.

Finally, in Fig. 9 we compare the results of our calculations for
carbon  WD   models  with   a  hydrogen   envelope  against   the
computations performed by Wood (1995). As we mentioned, the  mass
- radius relation have been the subject of many authors,  amongst
others,  Koester  1978,  Iben  \&  Tutukov  (1984), Mazzitelli \&
D'Antona (1986), Wood (1995). Here, we shall compare with  Wood's
models  since  they  have  been  thorougly  employed  by  the  WD
community. Note that the general  trend of our results and  those
of Wood is very similar. As Wood considered more massive hydrogen
envelopes  ($M_{\rm  H}/M_*=  10^{-4}$)  than  we  did,  we  have
recomputed models  with 0.6,  0.7 and  0.8 \msun\  and with  that
hydrogen  mass.  We  find  that,  for  the  same value of $M_{\rm
H}/M_*$,  Wood  models  have  radii  a  bit  larger than ours (of
course, for the case of $M_{\rm H}/M_*= 10^{-5}$ the  differences
are much larger).  Note that as  models evolve, such  differences
become smaller.
\begin{figure}[h]
\epsfxsize=8.8cm
\epsfbox{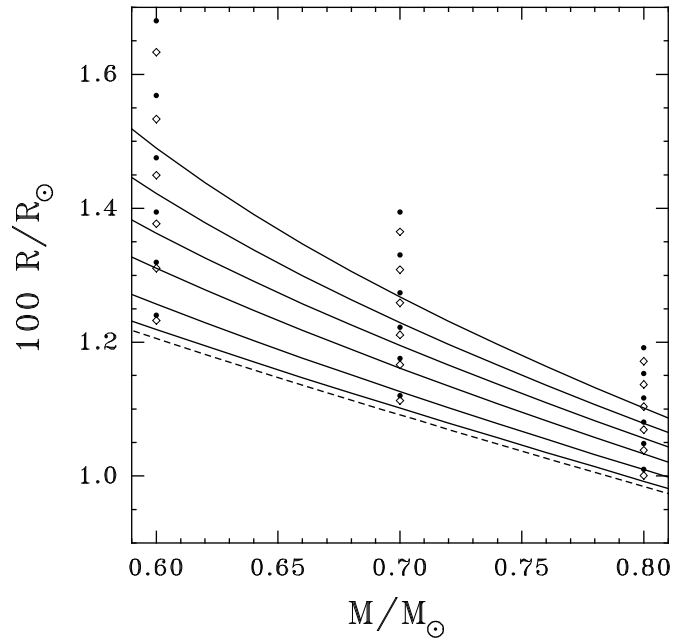}
\caption{
A comparison  of our results
with those of  Wood (1995). Solid  lines indicate the  previously
presented  results  for  carbon  core  models  including an outer
hydrogen envelope, solid  dots indicate the  results of Wood.  As
Wood considers more massive hydrogen layers ($M_H/M_*=  10^{-4}$)
than the one employed in most of this work, we have computed also
models  with  the  same  hydrogen  mass  envelope.  The   results
corresponding to these models are denoted with open
diamonds.}
\end{figure}

\section{Discussion and conclusions} \label{sec:conclusion}

In  this  work,  we  have  computed  accurate and detailed mass -
radius relations for white  dwarf (WD) stars with  different core
chemical compositions. In particular we have considered interiors
made up by helium, carbon, oxygen, silicon and iron surrounded by
a helium layer containing 1\% of the stellar mass. With regard to
the  presence  of  a  hydrogen  envelope,  we have considered two
extreme values:  $M_{\rm H}/M_*  = 10^{-5}$  ($M_{\rm H}/M_*  = 3
\times 10^{-4}$ for helium core models) and $M_{\rm H}/M_* =  0$.
The first three interior compositions are standard (according  to
stellar  evolution  theory),  whereas  iron  - rich interiors has
recently   been   suggested   on   the   basis  of  new  parallax
determinations for some objects (Provencal et al. 1998).

For  computing   each  sequence   we  employed   a  full  stellar
evolutionary  code  which  incorporates  most  of  the  currently
physical processes considered relevant to the physics of WDs.  We
computed a set of evolutionary sequences for each considered core
composition by  employing a  small step  in the  stellar mass. We
believe  that  these  calculations   may  be  valuable  for   the
interpretation of future observations of this type of WDs.

We have  also investigated  the effect  of gravitational settling
and chemical and thermal diffusion on low - mass helium WDs  with
envelopes made up  of a mixture  of hydrogen and  helium. To this
end, we included in our evolutionary code a set of routines which
solve the diffusion and heat flow equations for a  multicomponent
medium.  For  the  case  analysed  in  this  paper, we found that
diffusion gives  rise to  appreciable changes  in the theoretical
mass - radius relation, as compared with the case when  diffusion
is not considered (Driebe et al. 1998).

In Figs. 4 - 8 we included the data for 40 Eri B (\teff=16700 K),
EG  50  (\teff=21000  K),  Procyon  B  (\teff=8688  K) and GD 140
(\teff=21700 K) taken from Provencal  et al. (1998). In the  case
of 40 Eri  B for instance,  the observed mass,  radius and \teff\
are consistent with  models having a  carbon, oxygen and  silicon
interior and thin hydrogen envelopes. It is worth noting that the
observed determinations are also consistent with iron core models
with hydrogen envelope composition  but for models with  $\approx
55 000~  K$ (models  without hydrogen  envelope would  need to be
even  hotter).  Since  this  temperature  is  far larger than the
observed one we should discard an iron core for this object.

On the contrary, for the  cases of the other considered  objects,
they  fall  clearly  below  the  standard  composition sequences,
indicating a denser interior. If  we assume GD 140 and  Procyon B
to have an iron core, we find  that they fall on a sequence of  a
\teff\ compatible with the  observed value. Nevertheless, the  EG
50 mean radius is smaller than predicted for an iron core  object
for  the  observed  \teff.  Thus,  on  the  basis  of the current
observational determinations for EG 50, this WD seems to be  even
denser than an iron WD.

At present, it seems that the physics that determines the  radius
of a WD star is fairly well understood, thus the indication of an
iron core should not be expected  to be due to some error  in the
treatment  of   equation  of   state  of   a  degenerate  plasma.
Accordingly, if observations are confirmed to be accurate enough,
we should  seriously consider  some physical  process capable  to
produce an iron core for such low mass objects.

Detailed tabulations of the  results presented in this  paper are
available upon request from the authors at their e-mail address.

\section*{Acknowledgments}

O.G.B. wishes to acknowledge to Jan - Erik Solheim and the LOC of
the  11th  European  Workshop  on  White  Dwarfs held at Troms\/o
(Norway) for their  generous support that  allowed him to  attend
that meeting were  he became aware  of the observational  results
that  motivated  the  present  work.  We  also acknowledge to our
referee for his remarks and comments which significantly improved
the original version of this work.

%\begin{thebibliography}{99}

\end{document}